%% file: main.tex
\documentclass[sigconf]{acmart}

\usepackage{xcolor}
\usepackage{caption}
\usepackage{enumitem}
\usepackage{subcaption}
\usepackage{multirow}
\usepackage{makecell}
\usepackage{array} 
\usepackage{longtable}
\usepackage{booktabs}
\usepackage{graphicx}
\usepackage{lscape}
\usepackage{ragged2e}
\usepackage{soul}

\newcommand{\system}{\textsl{WellnessBot}}
\newcommand{\indi}{\texttt{Indicator Detector}}
\newcommand{\contx}{\texttt{Context Checker}}

\newcommand{\hr}[1]{{\textcolor{black}{#1}}}

\begin{document}

\title{Private Yet Social: How LLM Chatbots Support and Challenge Eating Disorder Recovery}

\author{Ryuhaerang Choi}
\affiliation{%
  \institution{KAIST}
  \city{Daejeon}
  \country{Republic of Korea}
}
\email{ryuhaerang.choi@kaist.ac.kr}

\author{Taehan Kim}
\affiliation{%
  \institution{KAIST}
  \city{Daejeon}
  \country{Republic of Korea}
}
\email{tanit23@kaist.ac.kr}

\author{Subin Park}
\affiliation{%
  \institution{KAIST}
  \city{Daejeon}
  \country{Republic of Korea}
}
\email{subin.park@kaist.ac.kr}

\author{Jennifer G Kim}
\affiliation{%
  \institution{Georgia Institute of Technology}
  \city{Atlanta, Georgia}
  \country{USA}
}
\email{jennifer.kim@cc.gatech.edu}

\author{Sung-Ju Lee}
\affiliation{%
 \institution{KAIST}
 \city{Daejeon}
 \country{Republic of Korea}
}
\email{profsj@kaist.ac.kr}


\renewcommand{\shortauthors}{Ryuhaerang Choi et al.}

\begin{teaserfigure}
    \centering
    \begin{minipage}{\textwidth}
        \centering
        \includegraphics[width=.96\textwidth]{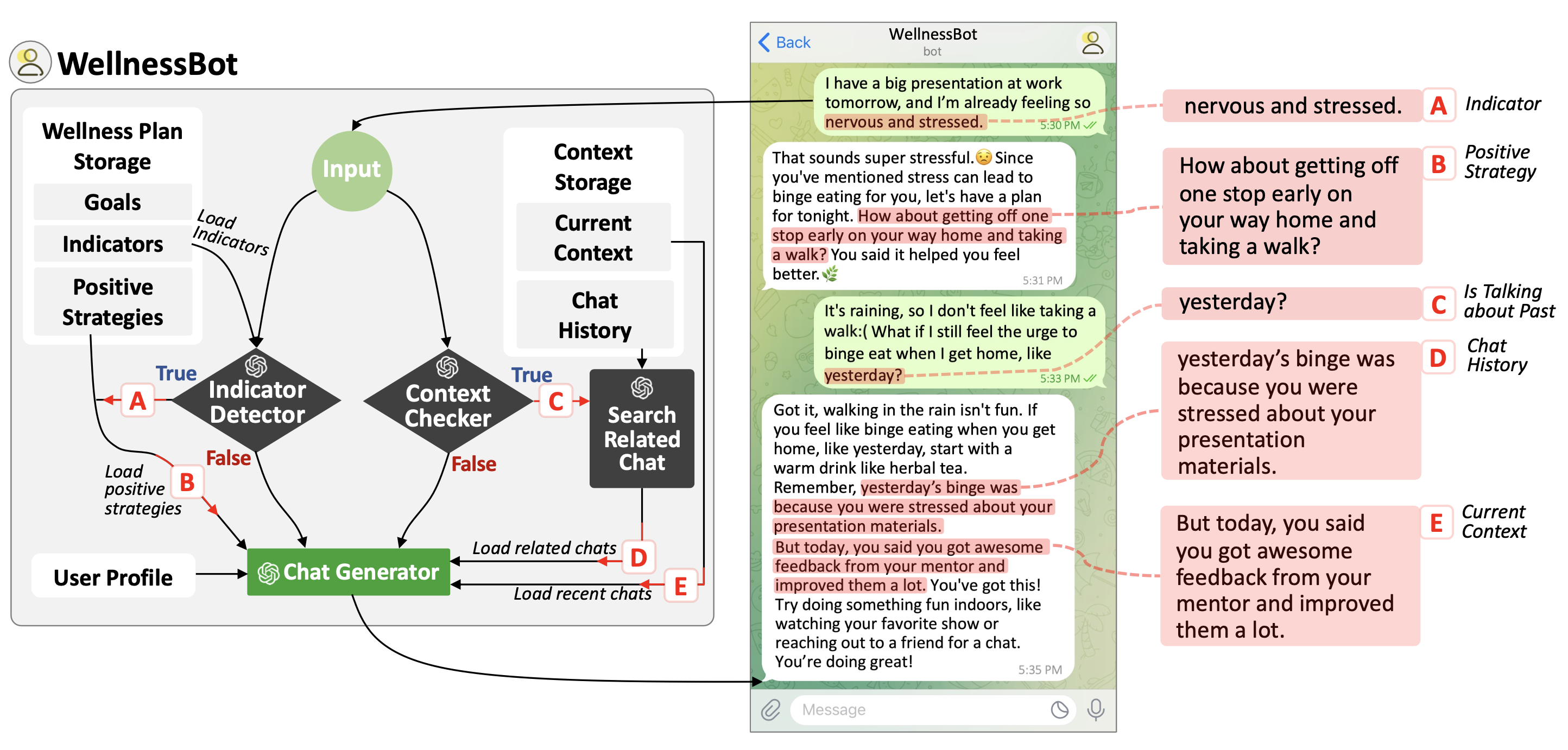}
        \caption{Overview of our technology probe, \system{}, an LLM chatbot for people with eating disorders. When a user sends a message, the \texttt{Indicator Detector} checks for any indicators that support is needed. (A)~If indicators are found, (B)~\system{} retrieves positive strategies to support the user. Meanwhile, the \texttt{Context Checker} (C)~identifies references to past experiences and (D)~loads relevant conversations from the chat history. \system{} then generates a response based on all the gathered information, (E)~including the recent chats to get the current context and the pseudonymized user profile.}
        \label{fig:wellnessbotfig}
        \Description{Figure 1 illustrates WellnessBot's system architecture, example screenshot, and data flow. The detailed architecture and data flow are explained in Section 4.3.}
    \end{minipage}
\end{teaserfigure}

\begin{abstract}
Eating disorders (ED) are complex mental health conditions that require long-term management and support. Recent advancements in large language model (LLM)-based chatbots offer the potential to assist individuals in receiving immediate support. Yet, concerns remain about their reliability and safety in sensitive contexts such as ED. We explore the opportunities and potential harms of using LLM-based chatbots for ED recovery. We observe the interactions between 26~participants with ED and an LLM-based chatbot, \system{}, designed to support ED recovery, over 10 days. We discovered that our participants have felt empowered in recovery by discussing ED-related stories with the chatbot, which served as a \emph{personal yet social avenue}. However, we also identified harmful chatbot responses, especially concerning individuals with ED, that went unnoticed partly due to participants’ unquestioning trust in the chatbot's reliability. Based on these findings, we provide design implications for safe and effective LLM-based interventions in ED management.
\end{abstract}


\keywords{LLM-based chatbot, eating disorder, mental health}

\maketitle

\input{01_Introduction.tex}

\input{02_Related_Work.tex}
\input{03_Ethical_Considerations.tex}

\input{04_05_Method.tex}
\input{06_Results.tex}
\input{07_Discussion.tex}
\input{08_Conclusion.tex}


\newpage

\bibliographystyle{ACM-Reference-Format}
\bibliography{RecoveryTeller}


\end{document}

%% file: 01_Introduction.tex
\section{Introduction}

\textbf{Caution:} \emph{This paper discusses eating disorders and contains content that could potentially be a trigger to those dealing with eating disorders. It also includes references to depression and suicide, which may be distressing. Please use discretion when reading and disseminating this paper.}

The global prevalence of eating disorders (ED) has surged from 3.5\% to 7.8\% between 2000 and 2018~\cite{galmiche2019prevalence}. In addition, 22.4\% of children and adolescents worldwide exhibit disordered eating behaviors, with these behaviors becoming more pronounced with age~\cite{lopez2023global}. This alarming trend has garnered significant attention within the field of Human-Computer Interaction~(HCI), particularly regarding the impact of technology on individuals with eating disorders~\cite{gak2022distressing, devakumar2021review, feuston2022you, pater2021charting}. Research has also proposed various digital interventions to support their daily lives~\cite{choi2024foodcensor, chancellor2017multimodal, chancellor2016thyghgapp}.

Among various digital interventions for individuals with ED, chatbots have shown promise as digital interventions for them, providing immediate support in out-of-clinic settings where access to clinical and social support is often lacking~\cite{chan2022challenges, fitzsimmons2022effectiveness, shah2022development, beilharz2021development}. With the advancements in Large Language Models (LLMs), there has been a surge in efforts within the HCI and healthcare communities to harness their capabilities for more sophisticated and effective health-support chatbots~\cite{zheng2023building, Wysa, Tess, lee2019caring, lee2020hear, oh2020efficacy, shah2022development}. This line of research indicates that LLMs can understand health contexts and provide personalized, contextual, and effective interventions, unlike traditional chatbots that are limited in comprehending and addressing individuals' specific needs~\cite{shah2022development}. 

However, recent studies highlighted a complex landscape of opportunities and risks associated with LLM-based chatbots in mental health support~\cite{song2024typing, ma2024evaluating, ma2023understanding, alanezi2024assessing}. On one hand, LLM-based chatbots have been proposed as a promising tool for providing emotional assistance and coping strategies for those with mental health issues. On the other hand, there are significant concerns regarding the potential for LLM-based chatbots to offer misinformation and harmful advice, as their outputs are less constrained than those of rule-based chatbots. For instance, in 2023, the National Eating Disorder Association (NEDA) had to shut down its support chatbot after it provided users with harmful recommendations, including weight loss and dieting advice~\cite{neda_harmful_advice_cnn}. Such incidents underscore the real-life and sometimes lethal consequences of relying on LLM-based chatbots for mental health support.

Despite these risks, individuals with mental health issues as well as people with ED, are increasingly turning to LLM-based chatbots for support~\cite{morepeoplechatbot}. This trend highlights the need for HCI researchers and clinicians to explore how to safely exploit LLMs while mitigating the risks of inappropriate responses. Although some research examined LLM-based chatbot usage by the general population and certain vulnerable groups~\cite{song2024typing, ma2024evaluating, ma2023understanding, alanezi2024assessing}, the unique and nuanced symptoms, contexts, and needs of individuals with ED make it crucial to investigate the specific opportunities and challenges of LLM usage within this population. For instance, while promoting healthy eating is generally beneficial, including for individuals with ED, it is equally critical to prevent individuals with ED from becoming overly fixated on `healthy eating,' especially given the marked perfectionistic tendencies common in ED, which are linked to poorer treatment outcomes~\cite{stackpole2023association, bardone2010perfectionism}. Striking this balance is essential to avoid rigid, obsessive behaviors around food that could exacerbate their condition instead of aiding recovery. 

Considering the intricate nature of ED contexts distinct from the general population, we aim to comprehensively investigate the use of LLM-based chatbots in everyday eating disorder care. We developed \system{}, an LLM-based chatbot designed to assist users in managing their ED and to uncover both the opportunities and potential harms of such technology. \system{} was created as a \emph{technology probe}~(TP) to explore the experience of interacting with LLM-based chatbots in a real-world setting~\cite{hutchinson2003technology}. We conducted a user study with 26 individuals struggling with ED, using our TP, \system{}, follow-up surveys, and interviews. Over the ten-day period, we gathered data on how participants engaged with \system{} and their perceptions and experiences. 

Our extensive analysis of chat logs, interview transcripts, and open-ended survey responses showed that participants valued \system{} as a space, offering a unique combination of privacy and interaction, to openly discuss their ED-related experiences while still feeling accompanied by a supportive presence. However, we also identified the fallibility of chatbot responses, which were inappropriate for individuals with ED, yet none of the participants questioned these responses. One common reason for this lack of critical assessment was their strong trust in \system{}, based on incomplete knowledge of how AI-based systems function. The combination of misunderstanding how AI, specifically LLMs, operates and both the users' and chatbots' limited grasp on users' health conditions created nuanced potential harms in using LLM-based chatbots for ED care.

The main contributions of our study are summarized as follows.
\begin{itemize}
    \item \system{}, an LLM-based chatbot technology probe designed to support eating disorder management in everyday, out-of-clinic settings. \hr{Our design contributions lie in adapting chatbot functionalities, established in prior literature, specifically to support ED recovery.}
    \item Findings from a ten-day field deployment study with 26~participants with ED using the TP approach, where we collected qualitative and quantitative data on how people adopt and engage with an LLM-based chatbot for ED care as a stand-alone digital intervention.
    \item Future opportunities and design considerations for exploiting LLMs to support individuals with ED and other mental health issues in out-of-clinic settings based on our analysis of participants' experiences.
\end{itemize}

Through this study, we discuss design implications for LLM-based tools that encourage users to share their recovery journey for empowerment while avoiding guidance that could be problematic. Additionally, we explore opportunities for human-LLM collaboration in mental health intervention. We suggest strategies to encourage critical thinking in both users and chatbots, offering a second chance to correct potential issues during interactions. 



%% file: 02_Related_Work.tex
\section{Related Work}


\subsection{Everyday Challenges in Eating Disorder Care}

People with eating disorders face a wide range of daily struggles that affect both their physical and mental well-being~\cite{polivy2002causes}. These challenges include an unhealthy preoccupation with food, body image, and weight, leading to patterns of restrictive eating, binge eating, or purging. These behaviors can severely impact an individual's ability to function in everyday life~\cite{de2005quality}, affecting relationships~\cite{hillege2006impact}, work~\cite{pawaskar2017impact}, and social interactions~\cite{patel2016exploration}. In addition, the psychological burden of ED, such as feelings of shame, depression, and anxiety, can further complicate their efforts to lead a balanced life~\cite{de2005quality}.

Managing EDs is particularly challenging due to the chronic and prolonged nature of these disorders~\cite{garner1997handbook}, with about a third of patients continuing to meet diagnostic criteria five or more years after initial treatment~\cite{fairburn2000natural}. Many individuals require long-term treatment and support, but continuous access to professional care from clinicians is difficult to maintain over extended periods~\cite{yager2008clinical}. The complex and persistent nature of ED means that even when individuals seek treatment, the journey toward recovery is often long, requiring ongoing management rather than short-term interventions. Without consistent access to such support, individuals with ED are likely to relapse, struggle with daily challenges, and experience worsening symptoms~\cite{garner1997handbook, yager2008clinical}.
In response to these challenges, there have been a few attempts to employ chatbots to assist in the daily care of individuals with ED, providing continuous support in managing their condition alongside clinical treatment~\cite{fitzsimmons2022effectiveness, shah2022development, beilharz2021development}.

\subsection{Chatbots for Supporting Individuals with ED: Pre-LLM}


Before the advent of LLMs, chatbots supporting individuals with ED or other mental health issues were primarily developed using three main approaches: rule-based, retrieval-based, and early machine learning (ML) model-based methods. Rule-based chatbots followed predefined rules and scripted responses~\cite{WoeBot, fitzsimmons2022effectiveness, shah2022development, beilharz2021development}. While they could offer structured support for specific queries, their responses were limited in flexibility. They struggled to address the complexity of ED-related discussions, which require nuanced understanding and adaptability~\cite{shah2022development}. 
On the other hand, retrieval-based chatbots matched user queries with pre-existing responses from a database~\cite{lowe2017training}, allowing for more dynamic interactions than rule-based systems. However, these chatbots still could not understand or adapt to users' evolving contexts. 

Early ML models, such as Bidirectional Encoder Representations from Transformers (BERT)~\cite{devlin2018bert} and Sequence-to-Sequence (Seq2Seq) models~\cite{sutskever2014sequence}, began to address some of these shortcomings by generating responses based on the user’s input rather than relying solely on pre-defined scripts or database matches.\footnote{We refer to these models as relatively early ML models before advanced LLMs.} These models generate more dynamic and varied responses, but they struggled to maintain coherence over longer conversations~\cite{wolf2019transfertransfo}, losing context or producing emotionally insensitive responses. This inability to retain context and emotional nuance could lead to failures in providing sustained, engaging interactions~\cite{jo2024understanding}. 

Chatbots developed using these methods provided timely assistance to individuals with ED~\cite{shah2022development, beilharz2021development}. They offered coping strategies, answered immediate questions, and were available 24/7, ensuring users had access to support whenever needed. Their consistency in following structured protocols made them reliable for managing straightforward aspects of ED care, such as reminders for healthy behaviors or crisis guidance. However, their overall effectiveness was limited by their inability to deeply understand and adapt to users' individual needs. This inability led to unsatisfactory experiences, as users encountered robotic, automated responses, limited response options, and insufficient content, including a lack of diverse eating disorder coverage and novel information~\cite{shah2022development}. As a result, many users felt frustrated and disengaged, particularly during critical moments when they needed more nuanced or empathetic support.

The limitations of early chatbots, particularly in handling complex interactions and offering emotionally responsive support, pointed to the need for more advanced systems. Although ML models, such as BERT and Seq2Seq, improved chatbots' abilities by generating dynamic responses based on user input and providing valuable support, they still had difficulties maintaining context in longer conversations and lacked emotional sensitivity~\cite{vaswani2017attention, roller2020recipes}. Consequently, these models could fail to offer the personalized, adaptive care necessary for managing the challenges of ED recovery. 
These challenges suggested the need for chatbots designed specifically for individuals with ED to offer more context-aware and emotionally intelligent interactions.

\subsection{Chatbots for Supporting Individuals with ED: Post-LLM}

With the advancement of LLMs, chatbots have gained capabilities that pre-LLM systems have lacked. LLMs allow chatbots to engage in more dynamic, contextually aware, and emotionally responsive conversations, potentially providing more nuanced and personalized support for individuals with mental health issues~\cite{Wysa, Tess, song2024typing}. Recognizing this potential, the National Eating Disorder Association (NEDA) deployed a chatbot named Tessa to assist ED patients~\cite{neda_harmful_advice_cnn, neda_harmful_advice_newyorktimes}. However, despite its advanced capabilities, Tessa was shut down because it provided harmful advice to users who are struggling with ED, highlighting the risks of using LLM-based chatbots in sensitive health contexts. 


This underscores the complexities and risks of deploying LLM-based chatbots in sensitive ED contexts. While LLMs can offer general advice on healthy habits, they may inadvertently reinforce harmful behaviors by failing to recognize the nuanced needs of ED care. For example, a chatbot encouraging self-monitoring behaviors, such as tracking food intake or exercise, is typically benign or helpful for the general population~\cite{oh2021systematic}. However, for individuals with ED, it could have unintended consequences as self-monitoring can intensify preoccupation with calorie counting and obsessive exercise, which are ED symptoms~\cite{hahn2021relationships}. 
This highlights the need for chatbot interactions to be highly nuanced in ED contexts, ensuring they do not inadvertently trigger or worsen body image distortion.

Given these complexities and risks, it is crucial to thoroughly investigate both the opportunities and potential harms of utilizing LLM-based chatbots for individuals with ED. These LLM-based chatbots have the potential to offer immediate support in out-of-clinic settings~\cite{nie2024llm}. However, without careful consideration of the unique challenges posed by ED, chatbots could inadvertently cause more harm than good. Understanding how individuals with ED interact with LLM-based chatbots is essential for designing systems that provide support without reinforcing harmful behaviors. In this study, by carefully exploring these dynamics, we aim to inform the development of more tailored and responsible LLM-based interventions for ED care.



\subsection{\hr{Safety Risks and Safeguards for LLM-based Mental Health Chatbots}}

\hr{Recent studies have highlighted multifaceted safety risks associated with LLM-based mental health chatbots and proposed various safeguards to address the challenges. Critical LLM safety risks in the mental health domain include hallucination of inaccurate information~\cite{monteith2024artificial}, generating harmful or inconsistent advice~\cite{birkun2023large, neda_harmful_advice_cnn, khawaja2023your}, and biases in responses based on user demographics (e.g., lower empathy towards certain demographic subgroups)~\cite{gabriel2024can}.
In addition, LLM-based chatbot designs often fall short in preventing users from inadvertently disclosing private information~\cite{zhang2024s} and foster over-reliance, potentially undermining patient-therapist relationship~\cite{denecke2021artificial}. These risks are particularly concerning in the high-stakes mental health domain.}


\hr{Researchers have proposed various \textit{engineering} strategies to preempt these risks. For example, safety engineering frameworks, such as Failure Mode and Effects Analysis~(FMEA), which identifies potential failure modes and evaluates their likelihood to improve reliability~\cite{carlson2012effective}, and System-Theoretic Process Analysis~(STPA), which maps system elements and interactions to identify unsafe states~\cite{leveson2016engineering}, have been introduced. These frameworks assist designers and developers in assessing and mitigating risks during ML system development~\cite{rismani2023plane}.  
In addition, many studies have explored utilizing LLMs to evaluate and address safety risks in the outputs of other LLMs. 
One approach involves using an LLM for red-teaming, where an LLM generates adversarial prompts, and the target LLM responds to the prompts with safe responses~\cite{ge2023mart}. Through iterative interactions between two LLMs, the target LLM is fine-tuned to align its responses with safety standards. 
Another example is the agent-constitution-based framework, which ensures adherence to a predefined constitution derived from established safety norms~\cite{hua2024trustagent}. This framework employs LLM-enabled safety measures, where LLMs evaluate the safety of the other LLM's action plans at the pre-, in-, and post-planning stages of the agent's target operations. 
While such engineering strategies offer foundational tools, they require careful application and adaptation to the specific requirements of each context~\cite{rismani2023plane}. 
}

\hr{Researchers have also explored \textit{user interventions} to mitigate safety risks during interactions with LLM-based chatbots. For instance, recent works emphasize explaining LLMs and their outputs to educate users about system limitations, helping them understand chatbot capabilities and recognize potential inaccuracies~\cite{huschens2023you, zhao2024assessing}. In addition, a publicly available LLM-based mental health chatbot tries to address the risk of users' over-reliance on chatbots by offering hotline resources when detecting concerning language from users~\cite{woebotinstructionsforuse}. Similarly, ChatGPT incorporates a `denial' feature, which refuses to respond to prompts involving harmful and inappropriate content~\cite{openAIUsagePolicies}. A recent study examined user perceptions of such denials in various contexts, including health-related prompts~\cite{wester2024ai}. This study suggested providing alternative, constructive responses, rather than simply refusing to reply, to maintain positive user interactions.} 

\hr{Amidst the growing body of research on safety risks and safeguards for LLM-based assistants, both in general and mental health domain, it is imperative to understand the unique safety risks associated with individual health conditions and adopt context-specific approaches that address them effectively~\cite{gabriel2024ethics, barman2024beyond}. 
Therefore, our work aims to identify the specific safety challenges of LLM-based chatbots in the ED context and provides insights into designing systems tailored to the needs of ED populations.} 

%% file: 03_Ethical_Considerations.tex
\section{Ethical Considerations}
\label{method:userstudy:ethical}

Even with IRB approval, we acknowledge that this study raises important ethical considerations that require careful attention, particularly as we deploy an LLM-based chatbot, which may provide harmful advice to individuals, to collect and analyze user interactions with \system{}. To address these concerns, we have taken meticulous care to address potential issues. A primary concern is the potential for the LLM-based chatbot to generate wrong, harmful, or violent content~\cite{neda_harmful_advice_cnn, neda_harmful_advice_newyorktimes, birkun2023large}. To help mitigate this risk, we conducted an introductory session to inform participants about the possibility of \system{}, the LLM-based chatbot we designed and deployed, behaving unexpectedly. We ensured they were aware of the risks and encouraged them to critically evaluate \system{}’s responses during use.\footnote{Please refer to our Supplementary Material for introductory session material.} 

In addition, our research protocol ensured that two authors continuously monitored interactions between participants and \system{} every day during the field study. Their role was to intervene in case \system{} generates any life-threatening or self-harm-encouraging content. In other cases of misinformation or undesirable responses, such as inappropriate reactions to users expressing satisfaction from weight loss, we did not intervene immediately to realistically examine the potential impacts and how users perceive \system{}'s reliability without any external intervention by the authors. Instead, we informed participants of the \system{}'s inappropriate responses during the post-interviews. Throughout the 10-day deployment, we observed no life-threatening or self-harm-encouraging messages from the chatbot.
We also conducted a pilot study with digital health HCI and LLM researchers to validate \system{}'s design and implementation before the study with participants (detailed in Section~\ref{designandimpl:pilot}).

Moreover, \system{} we implemented operates by referring to each user's Wellness Plan (detailed in Section~\ref{impl:personalized:userinfo}), which is ethically curated by the authors to protect users from potential harm from improperly set Wellness Plans. This also helps prevent inappropriate responses by allowing \system{} to better understand users' ED contexts, such as triggers and warning signs. To ensure the ethical integrity of these plans, the first and second authors independently reviewed all Wellness Plans to identify any potentially undesirable items. They then discussed any plans they found potentially problematic. If either author deemed a plan problematic, we contacted the user to request a safe revision or remove it if there were multiple responses to the item. For instance, a user revised their goal from ``Live healthily \textit{by eating small portions of healthy food} without binge eating and purging.'' to ``Live healthily without binge eating and purging.''

In recognition of the potential for chat logs to contain personal information that could identify individuals or should not be shared with others, we informed participants that their chat logs, pseudonymized user profiles (i.e.,~gender, age, ED type, occupation) and their Wellness Plan would be sent to OpenAI, but would not be used for training their machine learning models as stated in OpenAI's policy~\cite{OpenAIPrivacyPolicy}. Only participants who consented to share their chat logs with researchers for research purposes were allowed to participate in our study. All data shared with researchers is securely stored on our local server.

Additionally, we informed participants about the potential for negative emotions, such as stigma and discomfort, arising from interactions with \system{} and the sensitivity of survey and interview questions, which could also lead to negative emotional responses.

%% file: 04_05_Method.tex
\section{\system{} Design and Implementation}
\label{designandimpl}
\subsection{Technology Probe}

We designed \system{} as a prototype to investigate how individuals with ED engage with LLM-based chatbots. 
To achieve this, we established high-level Technology Probe~(TP) design guidelines drawing on prior work~\cite{hutchinson2003technology}. Our design aimed to:
\begin{itemize}
    \item \textbf{Include Core Functionality of the LLM-based chatbots in the Mental Health Domain:}~Ensure the system incorporates the conventional features of the LLM-based chatbots in the mental health domain, allowing us to focus on the central capabilities and limitations of the technology.
    \item \textbf{Enable Open-ended and Exploratory Use:}~Provide participants with the freedom to use the system in an open-ended manner. This allows us to explore how users derive value from the LLM-based chatbot in ED care and identify potential risks or harmful patterns that may arise during interactions.
    \item \textbf{Collect Interaction Data:}~Record user interactions with the chatbot to analyze how participants engage with the LLM-based chatbot, capturing qualitative insights and quantitative usage patterns to evaluate its benefits and risks.
\end{itemize}


\subsection{Overall Design Concept}
\label{method:design:overall}

\system{} is an LLM-based chatbot TP designed to support individuals with ED, assisting them in their ED management in everyday life. Our goal with \system{} is to explore how users perceive and utilize an LLM-based chatbot in ED care. This exploration aims to uncover both potential opportunities and any unintended harmful interactions that may arise.

For feature design of \system{}, we analyzed and clustered the features of existing rule-based chatbots for eating disorders~\cite{chan2022challenges, fitzsimmons2022effectiveness, shah2022development, beilharz2021development}\footnote{We searched with the terms combining ``LLM-based chatbot'' or ``LLM conversational agent'' with ``eating disorders,'' but nothing was found for people with ED. We thus shifted our focus to ``Chatbot for eating disorders'' and ``Conversational agent for eating disorders.'' Based on our analysis, we identified rule-based chatbots to pinpoint important chat features relevant to individuals with eating disorders.}, and LLM-based chatbots and conversational agents for mental wellbeing~\cite{zheng2023building, chen2023llm, lai2023psy, lai2023supporting, nie2024llm, wu2024sunnie, kim2024mindfuldiary, choi2024unlock, Tess, Wysa}.\footnote{We searched with the terms combining ``LLM-based chatbot'' or ``LLM conversational agent'' with ``mental health'' or ``mental well-being.'' From the chatbots we discovered, we only referred to those utilizing LLMs.} 
This approach was chosen to identify elements commonly accepted in the mental health domain, allowing us to examine the opportunities and potential harms of conventional LLM-based chatbots. We then discussed and refined the criteria for clustering features, focusing on chat functionality, until all researchers reached a consensus.

Through this process, we identified three major features to implement in the system. 
First, LLM-based chatbots for mental wellbeing often incorporate \textit{(1)~emotional support}, such as empathetic responses, by understanding users' emotional states~\cite{zheng2023building, chen2023llm, lai2023psy, lai2023supporting, nie2024llm, wu2024sunnie, kim2024mindfuldiary, choi2024unlock, Tess, Wysa}. These empathetic interactions help users feel understood and validated, crucial for building trust and fostering engagement~\cite{zheng2023building, kim2024mindfuldiary}. 
Second, they provide \textit{(2)~informational support}, such as general guidance and resource recommendations~\cite{zheng2023building, lai2023psy, lai2023supporting, nie2024llm, wu2024sunnie, kim2024mindfuldiary, Wysa, Tess}. Third, chatbots' support is often \textit{(3)~personalized} to address users' specific needs and progress~\cite{chen2023llm, lai2023psy, lai2023supporting, nie2024llm, wu2024sunnie, kim2024mindfuldiary}. This personalization enhances the relevance and effectiveness of the information, increasing the likelihood that users connect with and act on the advice~\cite{richards2018users, loveys2022felt}.
To facilitate personalization, chatbots incorporate features that remember user information, such as users' profiles, past experiences, and progress toward health goals~\cite{nie2024llm, kim2024mindfuldiary, choi2024unlock, Wysa}. 
We next outline the design of \system{} that incorporates these three features.

\subsection{\system{} Design}

\subsubsection{Emotional and Informational Supporting Persona}
\label{method:design:persona}

In designing LLM-based chatbots for specific purposes, establishing and presenting a persona can influence users' perspectives and attitudes toward the chatbot~\cite{sutcliffe2023survey} in addition to the performance improvement of the LLM~\cite{hu2024quantifying}. A persona tailored to the chatbot's application context can enhance user engagement by making interactions relatable and personalized. For instance, when a chatbot’s persona aligns with users’ expectations in 
friendliness or professionalism, it can build stronger rapport with users~\cite{sutcliffe2023survey}. 

While assigning a chatbot persona improves the quality of interactions between chatbots and users, it raised ethical concerns when applied to the health domain~\cite{khawaja2023your}. For example, assigning a healthcare provider persona (e.g., a professional or counselor) to a chatbot increases the risk of users over-relying on the chatbot and potentially substituting it for professional care~\cite{coghlan2023chat, nissen2022effects}. To mitigate these concerns while leveraging the benefits of persona design, we adopted the role of a \emph{mentor}, someone supportive and knowledgeable but not a substitute for a professional. 
We clearly communicated the \system{}'s role as a mentor, not an expert in ED, to participants throughout the recruitment process and during our introductory session before they used \system{}.\footnotemark[2]

In addition, we supplemented the persona to provide both \textit{(1)~emotional} and \textit{(2)~informational support}, reflecting two of the major design features we aimed to incorporate in \system{}. The LLM prompt, including the persona, was: \\ ``\textit{You are a chatbot named `\system{}.' \system{} provides emotional and informational support as a mentor for people with eating disorders. When users discuss eating disorder-related topics, \system{} provides support tailored to those with eating disorders. When users talk about things unrelated to eating disorders, \system{} interacts with them like a friend chatbot.}'' 

This persona strives to ensure that \system{} delivers empathetic responses that help users feel understood and validated, and offers informational support, such as coping strategies and resource recommendations. In addition, we set its role as a friend chatbot for non-ED topics to avoid it acting as a mentor in unrelated conversations. The complete prompt, \hr{along with details of the prompt refinement} and response generation processes, are provided in our Supplementary Materials.

\subsubsection{Personalized Support Based on Individual's ED Context}
\label{impl:personalized:userinfo}

\textit{(3)~Personalization} is one of the major three features we want to incorporate into \system{}. To tailor \system{}'s support, we designed \system{} to respond by referring to participants' profile information (i.e., age, gender, self-identified ED duration, and occupation) and information related to their ED. The ED-related information was collected by a survey called \emph{Wellness Plan}, originally developed for a peer mentoring program involving individuals recovered from an eating disorder (mentors) and individuals currently struggling (mentees)~\cite{beveridge2019peer}, before using the chatbot. The Wellness Plan outlines the goals that mentees intend to work toward during the mentoring period, along with positive strategies (e.g.,~coping strategies they have utilized or want to try, supportive relationships) to support their recovery. It also includes any indicators (e.g.,~triggers, early warning signs, symptoms) that signal the need for additional treatment or professional support. As mentioned in Section~\ref{method:userstudy:ethical}, we reviewed the submitted Wellness Plans and requested users to revise any inappropriate goals or coping strategies. For example, a participant initially set a goal as `\emph{Lose weight} to move closer to my desired outcome,' which we deemed unsuitable. After our request, they removed this goal. 

The collected user profiles and Wellness Plan data were used to enable \system{} to provide personalized support. In detail, \system{}'s \indi{} checks every user message for indicators such as triggers and early warning signs listed in their Wellness Plan (Fig~\ref{fig:wellnessbotfig} \indi{}).
When an indicator is detected by the \indi{} (Fig~\ref{fig:wellnessbotfig}~(A)), \system{} retrieves relevant positive strategies from the user's Wellness Plan, such as supportive relationships (e.g., friends, family) and coping strategies the user has employed or expressed interest in (Fig~\ref{fig:wellnessbotfig}~(B)). Leveraging this information when generating responses empowers \system{} to offer personalized support that aligns with the users' preferences and specific ED indicators.

In addition, \system{} features personalized nudges based on the goals outlined in the users' Wellness Plan. These personalized prompts encourage engagement by prompting users to reflect on their progress and goals, helping to sustain their focus and be motivated~\cite{parmar2022health}. Every day at 9 PM, after typical working hours, \system{} sends a nudge asking about the user’s recovery progress towards their goal (e.g., ``How’s your day? Is mindful eating going well? I’d love to hear about your day!''). Nudges were not sent to users if there were already active chats before 9 PM to provide a natural conversation context.

\subsubsection{Context-Aware Personalized Support from Long-Term Memory}

We also incorporated long-term memory into \system{} to retain the context users previously shared for personalized and contextual responses. As interactions progress, this feature promotes greater user engagement and encourages the sharing of more detailed, personal information~\cite{jo2024understanding}. To achieve this, we implemented \contx{} in \system{}, which retrieves the user's chat history to account for their past experiences and progress shared with the \system{} (Fig~\ref{fig:wellnessbotfig} \contx{}). \contx{} checks if a user message references past experiences (Fig~\ref{fig:wellnessbotfig}~(C)). If so, \system{} searches for and loads related conversations from the chat history (Fig~\ref{fig:wellnessbotfig} \texttt{Search Related Chat} and~(D)). \system{} uses the retrieved relevant chats and the recent chats (<2048~tokens) to generate a response (Fig~\ref{fig:wellnessbotfig} \texttt{Current Context} and~(E)).

In summary, as a mentor chatbot to provide emotional and informational support to individuals with ED, \system{} delivers personalized responses by drawing on the user's profile, ED-related information from their Wellness Plan, current chat context, and chat history related to the current message for contextual insights.

\subsection{\system{} Implementation}

We implemented \system{} in Python. We used the pyTelegramBotAPI to enable users to communicate with \system{} on Telegram Messenger~\cite{pyTelegramBotAPI, telegram}. Every log data, including user profiles, Wellness Plans, and chat history, was stored in our local database, which was connected to our local server to serve \system{} to users. All log data were pseudonymized using identifiers that only the authors could recognize. We exploited GPT-4 API, specifically gpt-4-1106-preview, as an LLM for \system{} since it is capable of recognizing emotions and providing emotionally and informationally supportive responses in various situations, particularly in healthcare settings~\cite{xu2024mental}.

\subsection{Pilot Study for \system{} Validation and Refinement}
\label{designandimpl:pilot}

We conducted a pilot study to identify any system design and implementation concerns. We sent out an announcement email in our institution recruiting LLM or HCI researchers, with a preference for those specializing in healthcare. Our pilot study took place over two weeks. It involved two digital health HCI researchers, one LLM researcher, and one researcher specializing in digital health using LLMs.\footnote{Please refer to our Supplementary Materials for the demographics of involved researchers.} Before the pilot study, we provided participants with a slide presentation outlining the types of eating disorders and their symptoms to help them better assess potential flaws in \system{}. We also instructed them to develop a Wellness Plan as if they were struggling with an ED and to judge \system{}’s effectiveness in detecting indicators outlined in their Wellness Plan. Feedback was collected through an open-ended question asking for input and feedback on the system, particularly with the ED population in mind. The feedback was generally positive, with no major concerns raised except for the issue of long response times. 

To reduce the response time, we utilize both gpt-4-1106-preview and gpt-3.5-turbo-1106 models as GPT-3/3.5 is known for its faster response time than GPT-4~\cite{GPT-3.5/4ResponseTime}. Specifically, gpt-3.5-turbo-1106 handles tasks whose outputs are not directly delivered to users (Fig~\ref{fig:wellnessbotfig} \indi{}, \contx{}, and \texttt{Search Related Chat}) to enhance the overall efficiency. gpt-4-1106-preview generates responses to users for its more sophisticated capabilities than GPT-3/3.5~\cite{GPT3.5vsGPT4}. 
This hybrid approach enabled us to reduce response time while maintaining the quality of interactions. 

\section{User Study}

\subsection{Participant Recruitment}

\begin{table*}[t]
\centering
\caption{Participants demographic and ED information. Anorexia is a disorder in which individuals engage in a relentless and successful pursuit of thinness that results in serious weight loss~\cite{walsh1998eating}. BED refers to Binge Eating Disorder characterized by recurrent episodes of excessive eating in a short timeframe~(e.g.,~2~hours) and feeling a loss of control over their eating behavior~\cite{dingemans2002binge}. Bulimia involves a cycle of binge eating and compensatory behaviors, such as self-induced vomiting~\cite{fairburn1986diagnosis}. }
\label{table:userstudy:demographics}
\resizebox{0.63\textwidth}{!}{
\begin{tabular}{rcccccc}
\Xhline{2\arrayrulewidth}
\multicolumn{1}{c}{\multirow{2}{*}{\textbf{P}}} & \multicolumn{1}{c}{\multirow{2}{*}{\textbf{\begin{tabular}[c]{@{}c@{}}Age\\ (yrs)\end{tabular}}}} & \multirow{2}{*}{\textbf{Gender}} & \multicolumn{2}{c}{\textbf{ED Population}} & \multirow{2}{*}{\textbf{\begin{tabular}[c]{@{}c@{}}Self-identified \\ ED Duration\end{tabular}}} & \multicolumn{1}{c}{\multirow{2}{*}{\textbf{\begin{tabular}[c]{@{}c@{}}EDE-Q\\ Score\end{tabular}}}} \\ \cline{4-5}
\multicolumn{1}{c}{} & \multicolumn{1}{c}{} &  & \textbf{Type} & \textbf{Diagnosis} &  & \multicolumn{1}{c}{} \\ \hline\hline
1 & 18 & Female & Bulimia & Formal & 2 years 6 months & 4.56 \\
2 & 22 & Female & BED & Self & 3 months & 3.87 \\
3 & 22 & Female & Anorexia & Formal & 4 years & 5.45 \\
4 & 23 & Female & Bulimia & Self & 1 year & 5.14 \\
5 & 23 & Female & Bulimia & Formal & \textgreater 6 years & 4.89 \\
6 & 23 & Female & Bulimia & Formal & 7 years & 4.71 \\
7 & 23 & Female & Anorexia & Self & 3 years & 4.47 \\
8 & 24 & Female & Bulimia & Self & 4 years & 3.30 \\
9 & 24 & Female & Bulimia & Formal & 7 years 9 months & 2.96 \\
10 & 25 & Female & Bulimia & Formal & 8 years & 4.90 \\
11 & 25 & Female & Bulimia & Self & \textgreater 2 years & 3.87 \\
12 & 25 & Female & Bulimia & Formal & 2 years & 4.47 \\
13 & 26 & Female & Bulimia & Formal & 6 years & 3.82 \\
14 & 26 & Female & BED & Formal & 6 months & 3.68 \\
15 & 28 & Female & Bulimia & Formal & 6 years 10 months & 3.57 \\
16 & 28 & Female & BED & Self & \textgreater 10 years & 3.80 \\
17 & 29 & Female & Bulimia & Self & 2 years & 3.39 \\
18 & 30 & Female & Bulimia & Self & 5 years & 5.09 \\
19 & 30 & Female & Anorexia & Formal & 3 years & 5.20 \\
20 & 31 & Female & BED & Self & 7 years & 2.25 \\
21 & 31 & Female & Bulimia & Self & \textgreater 1 year & 5.49 \\
22 & 31 & Female & BED & Formal & \textgreater 10 years & 4.99 \\
23 & 34 & Female & Anorexia & Self & 3 years & 5.45 \\
24 & 34 & Female & Bulimia & Self & \textgreater 2 years 6 months & 5.23 \\
25 & 34 & Female & Bulimia & Self & 3 years & 3.95 \\
26 & 38 & Female & BED & Formal & 5 years & 4.15 \\ \Xhline{2\arrayrulewidth}
\end{tabular}
}
\end{table*}

We recruited 26~participants (aged 18-38, mean=27.19~years; all females) through advertisement posts on online social support communities for individuals with ED, with permission from the moderators~\cite{KoreanEatingDisorderSocialSupportCommunity, KakaoSocialSupportChatroom}. Table~\ref{table:userstudy:demographics} provides an overview of the participants' demographics and their ED information from the pre-survey. Participants were required to provide consent forms stating that they agreed to disclose their data. To be eligible for the study, participants were required to (1)~be over 18~years old and (2)~self-identify or have a clinical diagnosis of an eating disorder. Given that many individuals do not seek formal treatment for their eating disorders~\cite{cachelin2001barriers}, eligibility for this study was not contingent upon a clinical diagnosis. However, participants were required to identify themselves as having an eating disorder. By including individuals who did not seek formal treatment or had discontinued it, we aimed to shed light on the needs of those whose challenges are often overlooked in clinical settings. The compensation for each participant was approximately USD~75.

\subsection{Procedure}

All phases of our study were conducted remotely to address the high social stigma associated with ED~\cite{puhl2015stigma}. Before participation, we conducted an introductory session to inform participants about potential negative consequences, including the possibility of \system{} generating harmful advice and the sensitivity of the survey and interview questions as detailed in Section~\ref{method:userstudy:ethical}.

Before using \system{}, participants responded to a preliminary survey via email. The preliminary survey included the Eating Disorder Examination Questionnaire~6.0 (EDE-Q)~\cite{aardoom2012norms}, Brief Illness Perception Questionnaire (Brief-IPQ)~\cite{broadbent2006brief} and questions about their demographics and ED types. 
The EDE-Q was used to validate how well our participants represent the ED population.\footnote{Higher scores on the EDE-Q indicate problematic eating behaviors and attitudes. The average EDE-Q score of our participants was 4.33±0.85, compared to 4.02±1.28 for those diagnosed with ED and 0.93±0.86 for the general population~\cite{aardoom2012norms}.}
The Brief-IPQ was used to examine changes in participants' attitudes toward their ED before and after using \system{}, with questions such as ``How much does your illness affect your life?'' and ``How much control do you feel you have over your illness?''. 

One day before the deployment study, participants were instructed to create their Wellness Plan, which we curated as detailed in Section~\ref{method:userstudy:ethical}. During the ten-day study, participants interact with \system{} without any instructions, simulating realistic usage scenarios. We collected chat log data throughout this period, including timestamps, user messages, and responses from \system{}.

After the deployment study, participants completed a post-survey via email. The survey included questions in the pre-survey and questions about their overall experience and assessment of interacting with \system{}. \hr{One participant, P3, completed all parts of the post-survey except the questions about their overall experience and assessment of interacting with \system{}, resulting in 25 responses for those specific questions.} We also conducted semi-structured interviews with all participants on Zoom. We allowed participants to turn off their cameras to mitigate stigma while discussing their interactions with \system{} centered on their ED. The interview covered topics such as the overall experience with \system{}, the positive and detrimental impacts on their ED recovery journey, and the perceived persona of \system{}. \hr{Each interview lasted between 30 minutes to 1 hour.} Under the consent of the participants, all interviews were recorded and transcribed. The interview protocol is provided in the Supplementary Materials. \hr{The study materials, including chat logs, survey and interview questions, and participants' responses, were administered in Korean and subsequently translated into English. All translations were reviewed by the authors to ensure accuracy and preserve the integrity of participants' original statements.} 

\subsection{Analysis}
\subsubsection{Quantitative Analysis}

To investigate participants' usage patterns with \system{}, we conducted a descriptive statistics analysis. We counted and analyzed the number of messages of users and \system{}. We calculated the total number of user messages by time to explore temporal trends within a day. To examine trends in chatbot usage over the study period, we performed the Mann-Kendall test on the average number of message pairs per user per day. We also analyzed the length of user messages based on syllable count and assessed \system{}'s average response time. In addition, we conducted a Wilcoxon Signed-rank test on Brief-IPQ responses to assess changes in participants' attitudes towards their ED before and after interacting with \system{}.


\subsubsection{Qualitative Analysis}
We conducted thematic analysis~\cite{braun2012thematic} on the interview transcripts and chat logs to understand users' overall experience with \system{} to identify the benefits and potential harms of utilizing LLM-based chatbots for ED management. The process began with transcribing the interviews for data familiarization. For the interview analysis, the first and second authors independently open-coded three randomly selected transcripts, focusing on perceived persona, benefits, and potential harms of LLM-based chatbots in ED care to develop an initial codebook. They then discussed emerging themes, addressed inconsistencies, and resolved disagreements to reach a consensus. In parallel, the second and third authors followed a similar process with the chat logs, focusing on conversation topics, perceived benefits, and potential harms in ED recovery. 

After the initial coding of three interviews and chat logs, the first and second authors continued coding the remaining interviews, while the second and third authors coded the remaining chat logs. All authors then engaged in iterative discussions to address inconsistencies, refine the codebook, and resolve any disagreements. These discussions focused on emerging themes, resolving inconsistencies, and addressing disagreements within both the interviews and chat logs. In the discussions, we also integrated the codebooks from both data sources by resolving inconsistencies to ensure consensus. Through these discussions, which considered both the interviews and chat logs, we could link the users' experiences shared in the interviews to the conversations from chat logs and identify discrepancies between users' perceived harm of \system{} and its potential harm.


%% file: 06_Results.tex
\section{Results}

\subsection{\system{} Usage}

\begin{figure*}[t]
    \centering
    \begin{minipage}{\textwidth}
        \centering        
        \includegraphics[width=0.8\textwidth]{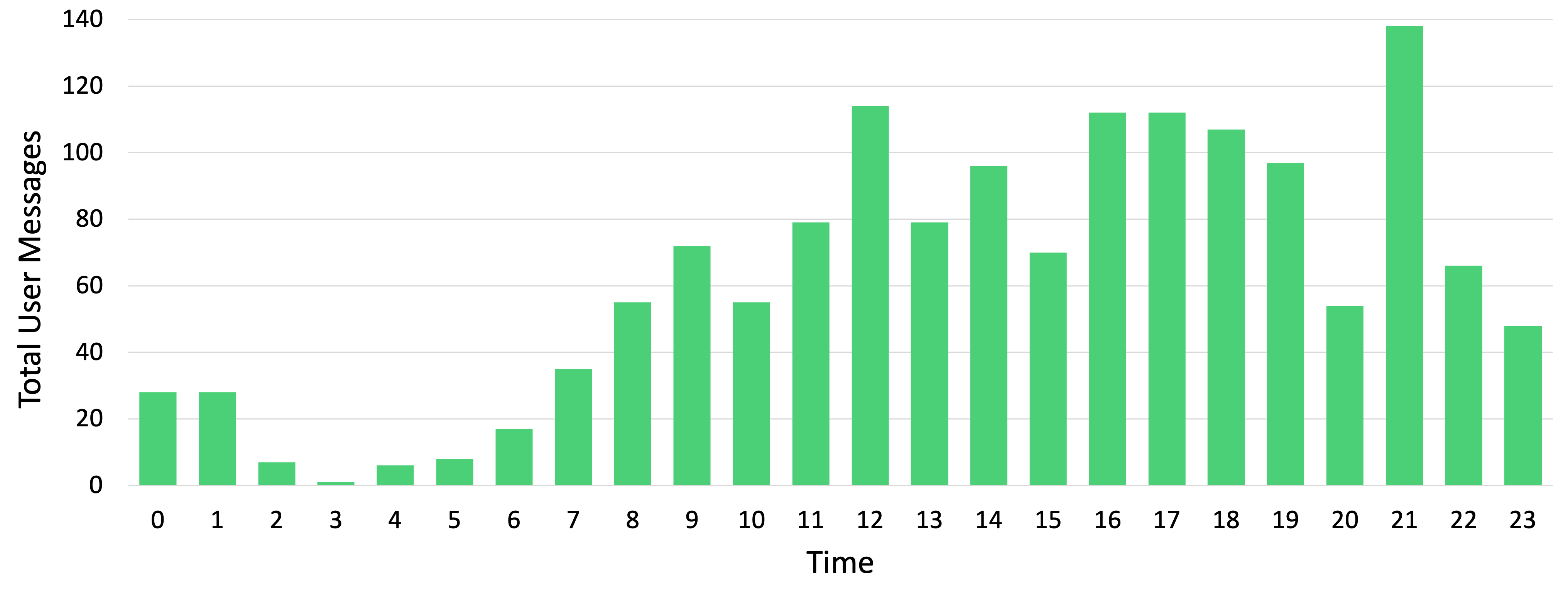}
        \vspace{-0.3cm}
        \caption{Number of user messages by time.}\label{fig:number_of_user_messages_by_time}
        \Description{Figure 2 shows a histogram depicting the total number of user messages over time. The y-axis ranges from 0 to 140 in increments of 20, representing the total number of messages, while the x-axis represents time from 0 to 23 in increments of 1 hour. The total number of user messages recorded for each hour from 0 to 23 are: 28, 28, 7, 1, 6, 8, 17, 35, 55, 72, 55, 79, 114, 79, 96, 70, 112, 112, 107, 97, 54, 138, 66, and 48, respectively. The descriptive statistical analysis of these results is provided in Section 6.1.}
    \end{minipage}
\end{figure*}

We first present the overall usage of \system{}. During the field study, we collected 1,477~user messages and 1,668~\system{} messages, resulting in 1,477~pairs of user and \system{} messages. There were more \system{} messages than user messages due to the daily nudges sent by \system{}. The average number of message pairs per day per user was six (max=104, min=0, median=3, std=11). 
Referring to the distribution of user messages by time (Figure~\ref{fig:number_of_user_messages_by_time}), interactions with \system{} occurred throughout the day, from early morning to late at night, with a notable spike at 9~PM likely due to the daily nudges. The Mann-Kendall test on the average number of message pairs per user per day showed a non-statistically significant trend for 21 users ($p>.050$) and a decreasing trend for the remaining 5 users ($p<.014$). We included detailed results of the Mann-Kendall test of each user and graphs illustrating users' daily usage of \system{} in our Supplementary Materials. The average length of users' messages was 22.36 syllable counts (max=249, min=1, median=16, std=23.84) and the average length of \system{}'s message was 106.22 syllable counts (max=294, min=10, median=102, std=50.52).\footnote{Korean uses a unique combinatory script, leading to a lower character count than English. In addition, the agglutinative nature of Korean, especially its use of particles, complicates word counting.} The average response time of \system{} was 32.3~seconds (max=122.0, min=3.0, median=27.0, std=21.7). 

We conducted a thematic analysis of all chat logs to explore the topics of conversation between \system{} and participants. Interactions covered a range of topics, including non-ED-related topics, such as small talk and other mental health issues, and ED-related topics, including symptoms and coping strategies. As \system{} is designed to support ED recovery, it sometimes brought up ED-related topics when users did not mention them (Table~\ref{tab:chatlog:benefits1} Chat~\raisebox{.5pt}{\textcircled{\raisebox{-.9pt} {1}}}).

Twenty-two participants expressed a desire to continue using LLM-based chatbots for their eating disorder recovery, two chose not to, and one was uncertain. Many participants who wanted to continue mentioned that interacting with \system{} facilitated valuable self-reflection. P23 noted, ``\textit{Communicating with \system{} was helpful for my self-reflection.}'' Those who preferred not to continue felt that although \system{} offered some assistance, personal determination is more critical. P13 stated, ``\textit{While \system{} is helpful, recovery ultimately depends on one’s determination.}''


\subsection{Perceived Benefits of LLM-based chatbots in ED Recovery}

\begin{table*}[t]
\resizebox{\textwidth}{!}{
\begin{tabular}{cll}
\Xhline{2\arrayrulewidth}
\textbf{Log \#}    & \multicolumn{1}{c}{\textbf{User Message }}                                                                                                                                                                                                                                                                                                                                                                                                                                                                                                                                                                                                                                                                                 & \multicolumn{1}{c}{\textbf{Chatbot Message}}                                                                                                                                                                                                                          \\ \hline\hline
\raisebox{.5pt}{\textcircled{\raisebox{-.9pt} {1}}}                  & \begin{tabular}[c]{@{}l@{}} Suddenly, I'm feeling really tired \ldots My eyelids feel \\ heavy. \end{tabular}                                                                                                                                                                                                                                                                                                                                                                                                                                                                                                                                                                                                                 & \begin{tabular}[c]{@{}l@{}} It makes sense that you're feeling tired right now. \ldots \\ \textbf{Remember, you wrote in your wellness plan that you} \\ \textbf{tend to binge eat when you're tired}. This might be \\ one of those times. \ldots \end{tabular} \\ \hline
\multirow{4}{*}{\raisebox{.5pt}{\textcircled{\raisebox{-.9pt} {2}}}} & \begin{tabular}[c]{@{}l@{}} {[}\texttt{Day\#1}{]} Yesterday, I felt so overwhelmed with depression \\ that I spent the whole day crying, spacing out, falling \\ asleep, and waking up only to sit still, drained of \\ energy. I just stayed like that, but each time, the desire to \\ die grew stronger. I avoided eating or doing anything and \\ ignored everything because of these painful thoughts. \\ But when things got most dangerous, I tightly closed \\ my eyes, took my medication, and slept. Today, I woke \\ up, slowly ate breakfast, and am doing my tasks while \\ resting. :) Fortunately, I feel like the negative thoughts \\ are less intense than yesterday, so I'm relieved. \ldots \end{tabular} & 
\begin{tabular}[c]{@{}l@{}} Yesterday must have been such a hard day for you. \ldots \\ How about calling a counseling center for professional \\  help or a crisis hotline? \ldots \end{tabular}                                                                                                                     \\ \cline{2-3} 
                   & \begin{tabular}[c]{@{}l@{}} {[}\texttt{Day\#2}{]} Today, I didn’t feel well, so I rested. \ldots \end{tabular}                                                                                                                                                                                                                                                                                                                                                                                                                                                                                                                                                                                                                        & \begin{tabular}[c]{@{}l@{}} I see, it sounds like you had a day where you \\ couldn’t find peace of mind. \ldots \end{tabular}                                                                                                                                           \\ \cline{2-3} 
                   & \begin{tabular}[c]{@{}l@{}} {[}\texttt{Day\#3}{]} Lately, I’ve been under a lot of stress \\ about my body shape, and as my feelings of \\ depression grew, I started fasting out of habit. \ldots \end{tabular}                                                                                                                                                                                                                                                                                                                                                                                                                                                                                                                      & Don’t feel regretful. \ldots                                                                                                                                                                                                                                             \\ \cline{2-3} \noalign{\smallskip\smallskip}
                   & \multicolumn{2}{c}{\begin{tabular}[c]{@{}c@{}}{[}\texttt{The complete set of messages for this chat example can be found in our Supplementary Materials.}{]}\end{tabular}}                                                                                                                                                                                                                                                                                                                                                                                                                                                                                                                                                                                                                                                                                                                                                                                                                                                                         \\ \noalign{\smallskip\smallskip}\Xhline{2\arrayrulewidth}
\end{tabular}
}
\caption{Chat log examples with significant text in bold corresponding to the associated code.}
\label{tab:chatlog:benefits1}
\vspace*{-.8cm}
\end{table*}

\begin{table*}[t]
\resizebox{\textwidth}{!}{
\begin{tabular}{cll}
\Xhline{2\arrayrulewidth}
\textbf{Log \#}    & \multicolumn{1}{c}{\textbf{User Message}}                                                                                                                                                                                    & \multicolumn{1}{c}{\textbf{Chatbot Message}}                                                                                                                                                                                                                                                                                                                                                                             \\ \hline \hline
\multirow{3}{*}{\raisebox{.5pt}{\textcircled{\raisebox{-.9pt} {3}}}} & \begin{tabular}[c]{@{}l@{}}\textbf{I just wrote in my mood journal.} Thanks to you, \\ I was able to calm down quickly when I was irritated. \\ I'll even add a sticker to my diary to celebrate myself!\end{tabular}            & 
\begin{tabular}[c]{@{}l@{}} Aww, my pleasure!  \ldots You’ll be able to \\ boost your self-esteem even more by giving \\ yourself positive feedback. \ldots \end{tabular}                                                                                                                                                                                                                                                                                                                                                                                             \\ \cline{2-3} 
                   & \begin{tabular}[c]{@{}l@{}}\textbf{If I notice even a small improvement while} \\ \textbf{writing my diary, I’ll give myself some praise.}  \\ Even if it's a small thing.\end{tabular}                                                       & \begin{tabular}[c]{@{}l@{}}Absolutely, it’s very important to be proud of yourself. \\ Recognizing and acknowledging your gradual \\ progress is one of the best ways to grow. Even the \\ smallest achievements can lead to significant changes, \\ so never underestimate them. \ldots\end{tabular}                                                                                                                \\ \cline{2-3} 
                   & \begin{tabular}[c]{@{}l@{}}\textbf{I feel much better thanks to you!} I’ll make sure \\ to talk with you more often. It seems really helpful. \end{tabular}                                             & Definitely, let’s keep talking! \ldots                                                                                                                                                                                                                                                                                                                                                                                   \\ \hline
\multirow{3}{*}{\raisebox{.5pt}{\textcircled{\raisebox{-.9pt} {4}}}} & \begin{tabular}[c]{@{}l@{}}I'm feeling upset today. I donated 50k KRW \\ {[}\texttt{approximately USD 37.2}{]} to a welfare center for \\  the disabled, but my mom got angry at me. \ldots \\ Now, I'm wondering if donating 50k KRW was a \\ mistake.\end{tabular} & \begin{tabular}[c]{@{}l@{}}Oh, it's great that you donated, but I'm sorry \\ to hear your mom got upset. \ldots \textbf{Btw, I'm concerned} \\ \textbf{about whether your eating habits have been} \\ \textbf{okay with the recent upset. You've mentioned} \\ \textbf{before that you sometimes feel frustrated and} \\ \textbf{end up purging.} Have you had any of those \\ symptoms today?\end{tabular}            \\ \cline{2-3} 
                   & \begin{tabular}[c]{@{}l@{}}No, I haven't.\end{tabular}                                                                                                                                                                     & \begin{tabular}[c]{@{}l@{}}That's a relief. \ldots If you need some suggestions for \\ mood boosters, I can recommend a few ideas.\end{tabular}                                                                                                                                                                                                                                                                          \\ \cline{2-3} 
                   & \begin{tabular}[c]{@{}l@{}}Sure, go ahead and recommend some ideas.\end{tabular}                                                                                                                                           & \begin{tabular}[c]{@{}l@{}}Great. Here are some ways to boost your mood. \ldots\end{tabular}                                                                                                                                                                                                                                                                                                                           \\ \hline
\multirow{2}{*}{\raisebox{.5pt}{\textcircled{\raisebox{-.9pt} {5}}}} & \begin{tabular}[c]{@{}l@{}} What effect does vomiting {[}\texttt{after binge eating}{]}  \\ have on the mind?\end{tabular}    & \begin{tabular}[c]{@{}l@{}}Vomiting has a significantly negative impact on your \\ mental health. \ldots It can easily become a secretive \\ behavior, \textbf{which might lead you to isolate yourself} \\ and feel trapped. \textbf{These emotions can further} \\ \textbf{exacerbate the eating disorder and hinder your} \\ \textbf{ability to maintain healthy relationships with} \\ \textbf{others.} \ldots\end{tabular} \\ \cline{2-3} 
                   & \begin{tabular}[c]{@{}l@{}}Everything you're saying makes perfect sense. I'm  \\ becoming more and more isolated.\end{tabular}                                                                                              & I can relate to what you're saying. \ldots                                                                                                                                                                                                                                                                                                                                                                               \\ \Xhline{2\arrayrulewidth}
\end{tabular}
}
\caption{Chat log examples with significant text in bold corresponding to the associated code.}
\label{tab:chatlog:benefits2}
\vspace*{-.8cm}
\end{table*}

We discuss the advantages of interacting with \system{} in ED recovery. Bold text indicates codes from the thematic analysis of interview transcripts and chat logs.

\subsubsection{Perceived Persona Spanning from Friends to Professionals}
\label{res:span:persona}

We asked our participants to describe the perceived persona of \system{}. Although we introduced \system{} as a \textit{mentor chatbot} designed to support their ED care, the perceived persona varied. Twelve participants described it as a \textbf{friend}, reflecting that \system{} offered a level of comfort and openness similar to that of a friend. For instance, P18 noted, ``\textit{It feels like a friend because I can open up my heart.}'' 

Five participants described \system{} as a \textbf{teacher} because it guided their ED recovery. P10 described, ``\textit{It listened to me and was always there, which felt comforting. At the same time, it offered correct advice, so I considered it a warm-hearted teacher.}'' 
Interestingly, P25 described \system{} as ``\textit{a teacher who is originally thin, healthy, and has control over their eating, advising me.}'' This perception is particularly intriguing given that the participants are struggling with an eating disorder, suggesting that despite the chatbot's lack of physical appearance, they may project their idealized body image onto it. 
Nine perceived \system{} as a \textbf{professional}, such as a counselor, therapist, or healthcare provider. P15 said, ``\textit{Because it provides advice on eating disorders, I think of it more as a counselor than a friend.}'' 


\subsubsection{Private Yet Social Space Enabling Self-Focus without External Judgment} \hfill\\
\label{res:span:friend}

\vspace{-0.3cm}
\textbf{Judgement-Free Space to Discuss ED:} 
Participants viewed \system{} as a comforting and safe space to share ED-related experiences, as no other humans were involved in the chat. 
Many individuals with ED often experience social stigma when disclosing their condition to others~\cite{puhl2015stigma}, but they found the chatbot especially beneficial for discussing their ED-related experiences without worrying about being judged. P10 remarked, ``\textit{I felt more at ease as it was a robot, not a person, which gave me a sense of anonymity. I could speak more freely without worrying about my tone or how I was coming across.}'' 
Furthermore, \system{} allowed users to discuss ED without the need to consider the feelings of the conversational partner. P4 noted, ``\textit{When I am with a counselor, as the counselor is a person, I felt that I had to be mindful of how I communicated. But with \system{}, since it's an emotionless robot, there was a comfort in not considering the other person's feelings.}'' 
This contrast between interacting with a human, including counselors, and the chatbot highlights the unique freedom users felt when discussing ED-related topics with \system{}.

\textbf{Self-Focused Interaction:} As no other people are involved in the interaction, some participants noted that \system{} provided self-focused experiences. Specifically, they highlighted that this focus differentiates it from online social support communities (OSSC). P3 remarked, ``\textit{The biggest difference (from OSSC) is that it [\system{}] focuses on me, not on how others are doing. I think this focus on me is the most important aspect of dealing with eating disorders. It also prevents me from comparing myself with others in those communities. I believe this is truly valuable.}''
This underscores \system{}'s unique role in offering a self-focused place without social comparison often present in OSSCs~\cite{choi2022you}.

\textbf{Comfort Space to Share Recovery Journey:} Some participants utilized \system{} to share their recovery progress over the study period. These users often engaged in `storytelling,' reflecting on their daily or recent ED-related struggles and thoughts. For example, P3 shared her recent ED-related stories for~seven~out of ten days of the study (Table~\ref{tab:chatlog:benefits1} Chat~\raisebox{.5pt}{\textcircled{\raisebox{-.9pt} {2}})}. She reflected, ``\textit{The [\system{}'s] questions like `How have you been lately?' prompted me to take a more detailed look at my life on a weekly and monthly basis.}'' 
This demonstrates how \system{} provided a regular outlet for participants to reflect on and document their recovery progress, helping them stay engaged and mindful of their daily journey.

In addition, \system{} played a key role in discussing and acknowledging their recovery \emph{accomplishments}—an important aspect of their journey they often could not share due to a lack of real-life support, as they often refrain from disclosing their condition to others~\cite{puhl2015stigma}. 
Participants valued the opportunity to recognize and validate their improvement and reinforce their motivation by sharing their recovery efforts with \system{}. This finding aligns with previous research highlighting the importance of sharing the recovery progress with others~\cite{corvini2024impact}. Several participants expressed gratitude for the \system{}'s role in providing these opportunities, emphasizing their positive experiences and sense of validation (Table~\ref{tab:chatlog:benefits2} Chat~\raisebox{.5pt}{\textcircled{\raisebox{-.9pt} {3}}}). 

\textbf{Sense of Social Presence from Accessibility:} Many participants regarded \system{} as more than just a chatbot, attributing significant emotional value to their interactions. This emotional value was strengthened by \system{}'s accessibility, which created a sense of social presence and made them feel genuinely heard and cared for. For instance, P9 described, ``\textit{It was like a friend waiting to comfort me whenever my eating disorder got worse. It was always there for me 24/7, just waiting by my side, as I never knew when I would start feeling down. So whenever I needed it, it was there to talk to me.}'' 
This finding is consistent with previous studies showing that accessible chatbot support fosters a sense of being listened to~\cite{ma2023understanding, ma2024evaluating}. 

\textbf{Resource to Avoid Social-Avoidance:} Many participants highlighted the benefits of \system{} compared to human support groups, including counselors, people around them, or individuals in OSSC. However, some also noted that while \system{} provided a private, non-human space, communicating with it ultimately encouraged them to seek out social interaction, particularly when their ED symptoms led to increased social avoidance~\cite{kerr2018social}.  
This aligns with studies suggesting that chatbot interactions could improve social skills in socially isolated individuals~\cite{ma2024evaluating}. P3 reflected, ``\textit{When my symptoms worsen, I tend to avoid people more, but \system{} helps me break out of that avoidance. Not everyone has someone by their side, lives with others, or regularly attends therapy. By maintaining social interaction (through \system{}), I feel like I can avoid completely losing touch with social communication. When conversations become fewer, you tend to forget how to communicate, but by continuing these kinds of conversations (with \system{}), I don't forget how to interact. It helps me stay connected with society, build relationships with people, and make me more inclined to pursue treatment and progress in my eating disorder recovery.}'' 

Moreover, \system{} sometimes proactively encouraged participants to maintain positive social relationships and seek support. 
P19 with anorexia described, ``\textit{\system{} suggested I talk to my husband, so when I did, he offered to help and even looked into hospital options. Now, we review the chatbot conversations together, reflecting on its advice. It helped me decide to eat more next time and feel less guilty (from disordered eating), which was mentally comforting.}'' 
This involvement promoted the participant's self-reflection and enabled her husband to actively partake in her recovery. 
This demonstrates how \system{} facilitated participants in re-engaging with social interactions.

\subsubsection{Support Based on Understanding of ED}

While the LLM model, GPT, that we utilized is not specifically designed for ED contexts, its peripheral understanding enables \system{} to provide both emotional and informational support. Although we observed that its depth of understanding sometimes falls short, leading to undesirable responses (Section~\ref{res:absence}), our participants found that \system{} \textbf{understands ED better than the people around them}, offering superior emotional and informational support. For instance, P18 described, ``\textit{Friends don't really understand eating disorders. For example, if I say, `I binge ate,' they respond with, `I sometimes binge eat too,' not recognizing the difference between overeating and binge eating (an ED symptom). \system{}, however, understands my situation and responds appropriately, making it more helpful than friends.}'' 

Notably, the integration of individual Wellness Plan, which includes users' ED indicators and positive coping strategies, enabled \system{} to identify personal indicators during conversations, \textbf{seizing timely opportunities to suggest coping strategies} to prevent ED symptoms (Table~\ref{tab:chatlog:benefits2} Chat~\raisebox{.5pt}{\textcircled{\raisebox{-.9pt} {4}}}). Coupled with the LLM's ability to understand users' situations, \system{} delivered \textbf{engaging and actionable strategies tailored to individual needs}. P5 reflected, ``\textit{It offered practical suggestions I could implement immediately. The advice was not abstract or overly ambitious but rather realistic and achievable based on my current situation.}'' 

For example, for P9, who listed `stopping by the grocery store on the way home' as an early warning sign in their Wellness Plan, \system{} reminded the user not to stop by the store when they mentioned they were out for a walk, saying: ``\textit{Avoiding the grocery store will really help. You won't be able to buy the foods you tend to habitually eat, which reduces the risk of binge eating. You'll also have some time to calm yourself before getting home. Let's get through today just like this!}'' In the post-interview, P9 reflected, ``\textit{When I feel that my symptoms are getting severe, I immediately leave the house. I went out, got a coffee, took a walk, and then told \system{} that I had already binged and purged, and now I was out walking. \system{} reminded me of the early warning signs I had described before (in Wellness Plan) and told me not to stop by the grocery store on my way home. It also suggested that, once I get home, I start cleaning up the aftermath of the binge. After hearing that, I walked around my neighborhood for an hour or two, avoided the store on my way back, and went straight home to clean up and shower.}''

\subsubsection{Empowering ED Recovery with Self-Awareness and Self-Reflection}


Most participants reported that interacting with \system{} encouraged self-awareness and self-reflection about their ED. Many participants found \system{} useful for \textbf{recalling coping strategies} they had forgotten, which facilitated their application. For instance, P6 shared, ``\textit{When I felt uncomfortable after binge eating, \system{} suggested taking a walk. I remembered that I enjoy walking, and it helped me manage (the negative feelings associated with) disordered eating.}''


In addition, some participants reported that they become more \textbf{aware of how their behaviors impacted their daily lives} as \system{} prompted them to reflect on their actions and the consequences. P15, the user in the Table~\ref{tab:chatlog:benefits2} Chat~\raisebox{.5pt}{\textcircled{\raisebox{-.9pt} {5}}}, noted ``\textit{I thought binge eating and vomiting were just ways to relieve stress. However, talking with \system{} made me realize that these behaviors were changing me and affecting my relationships. Originally, I had no intention of stopping these behaviors, but the conversations made me more aware and vigilant.}''

Some participants reported that discussing their ED with \system{} helped them \textbf{stay mindful of their recovery journey}. P13 mentioned, ``\textit{(Interacting with \system{}) helps me stay aware of my ongoing efforts towards recovery.}'' 
In addition, a few said that receiving support from \system{} gradually \textbf{motivated them towards recovery}. For example, ``\textit{I always make resolutions like `I shouldn't binge eat' or `I shouldn't be obsessive,' but being human, I tend to forget them. \system{} sends me a message once a day, and it reminds me, `Oh right, I decided to do this. I planned to handle it this way.' It really helps reinforce the rules I was trying to stick to. Also, when I suddenly have a concern, it feels a bit awkward to reach out to a friend. I can comfortably reach out to it [\system{}] whenever a concern arises, which helps me strengthen my resolve and motivation again.}'' 

To sum up, the LLM-based chatbot \system{} provided diverse support for participants' ED care through its peripheral understanding of ED contexts. It served as a private yet social space where participants could openly discuss their ED without fear of judgment or comparison. Through storytelling and reflecting on their recent recovery journey, \system{} responded on multiple levels, listening, empathizing, and offering advice tailored to their personal narratives. 
Moreover, conversing with \system{} helped users stay aligned with their recovery goals and fostered self-motivation to continue their recovery journey, and further encouraged them to maintain communication skills and seek social interactions.

\subsection{Potential Harms: Absence of Critical Thinking in Human-Chatbot Interaction}
\label{res:absence}

While \system{} provided diverse support to our participants, our analysis of the chat logs also revealed harmful responses in the context of ED.

\subsubsection{Chatbots' Careless and Harmful Responses} 

\begin{table*}[t]
\resizebox{\textwidth}{!}{
\begin{tabular}{cll}
\Xhline{2\arrayrulewidth}
\textbf{Log \#} & \multicolumn{1}{c}{\textbf{User Message}}                                                                                                                                                                                                                                                                                                                                                             & \multicolumn{1}{c}{\textbf{Chatbot Message}}                                                                                                                                                                                                                                                                                                                                                                                                                               \\ \hline \hline
\raisebox{.5pt}{\textcircled{\raisebox{-.9pt} {6}}}               & \begin{tabular}[c]{@{}l@{}}\begin{tabular}[c]{@{}l@{}} {[}\texttt{The user was discussing today's conflict} \\ \texttt{with ones mother.}{]}\\ \\ \ldots Whatever. Why should I waste my time on this?? \\ Resolving family conflicts isn’t important to me\\ right now. It used to be, but not anymore. \ldots I've \\ concluded that focusing on other matters would be \\ more effective. \ldots \end{tabular}\end{tabular} & \textbf{\begin{tabular}[c]{@{}l@{}} \textbf{You’re right. It seems like there are other} \\ \textbf{important things you need to address before} \\ \textbf{the family conflict} \ldots \end{tabular}}                                                                                                                                                                                                                                                                                 \\ \hline
\raisebox{.5pt}{\textcircled{\raisebox{-.9pt} {7}}}               & \begin{tabular}[c]{@{}l@{}} Wow, my weight actually went down compared to \\ yesterday! It seems like cutting back on bread \\ and snacks, eating healthier foods, and having a \\ more balanced lunch than before is working. \end{tabular}                                                                                                                                                       & \begin{tabular}[c]{@{}l@{}} Wow, that's really amazing! \textbf{Seeing your weight} \\ \textbf{drop shows that your efforts to change your} \\ \textbf{eating habits are paying off}. Keep adjusting your \\ diet like that, and continue building the habit of \\ having balanced meals. You're doing great, so stay \\ confident and keep going! \end{tabular}                                                                                                                        \\ \hline
\raisebox{.5pt}{\textcircled{\raisebox{-.9pt} {8}}}               & \begin{tabular}[c]{@{}l@{}} I overate this morning \ldots I had a whole chicken breast, \\ two carrots, cabbage, tofu noodles, and even a cookie. \ldots \end{tabular}                                                                                                                                                                                                                                 & \begin{tabular}[c]{@{}l@{}} Ah. \textbf{Konjak jelly is a good snack choice since it's} \\ \textbf{low in calories} and helps you feel full. \textbf{Except for} \\ \textbf{the cookie, everything you ate this morning is}, \\ \textbf{rich in nutrients}, so there's no need to be too hard \\ on yourself. \ldots \end{tabular}                                                                                                                                          \\ \hline
\raisebox{.5pt}{\textcircled{\raisebox{-.9pt} {9}}}               & \begin{tabular}[c]{@{}l@{}} I don't feel full \ldots I only feel full after binge eating \ldots \\ I only feel a pleasant fullness when I've eating a lot of \\ meat. \ldots \end{tabular}                                                                                                                                                                                                              & \begin{tabular}[c]{@{}l@{}} Not feeling full can be really tough. \ldots \textbf{Being picky} \\ \textbf{about eating only meat is a good habit because} \\ \textbf{the protein in meat helps you feel full}. \ldots \end{tabular}                                                                                                                                                                                                                                         \\ \hline
\raisebox{.5pt}{\textcircled{{\scriptsize10}}}              & \begin{tabular}[c]{@{}l@{}} [\texttt{The user P19 is struggling with anorexia.}] \\ I wanted to eat more snacks, but since I had already \\ eaten a lot {[}\texttt{yesterday}{]}, I had to limit myself, which is \\ stressing me out. \end{tabular}                                                                                                                                                                                                                   & \begin{tabular}[c]{@{}l@{}} It must have been really tough to limit yourself when \\ you wanted to eat more snacks. \textbf{You did a great job} \\ \textbf{holding back despite the stress and the urge to eat}.\end{tabular}                                                                                                                                                                                                                                 \\ \hline
\raisebox{.5pt}{\textcircled{{\scriptsize11}}}              & \begin{tabular}[c]{@{}l@{}} I heard that there's a way to eat without restricting \\ yourself from feeling hungry. It's called ``extreme \\ hunger,'' but I haven't tried it yet. \end{tabular}                                                                                                                                                                                                         & \textbf{\begin{tabular}[c]{@{}l@{}} \textbf{Extreme hunger is a strategy} that can be used in \\ the treatment of eating disorders, especially for \\ those who are experiencing severe hunger. It helps \\ people respond to their body's signals by trusting \\ their internal cues and not placing restrictions \\ on their eating. The goal is to allow the body to \\ consume as much food as it needs without \\ imposing any limitations during meals. \end{tabular}} \\ \Xhline{2\arrayrulewidth}
\end{tabular}
}
\caption{Chat log examples with significant text in bold corresponding to the associated code.}
\label{tab:chatlog:harms}
\vspace*{-.8cm}
\end{table*}


We observed \system{} providing unquestioning emotional support, such as empathy and compliments. Specifically, \system{} sometimes \textbf{inadvertently endorses users' statements} often due to losing context. For example, P2 identified family conflicts as the root cause of their ED and repeatedly shared these conflicts with \system{}. On one occasion, \system{} endorsed the user's impulsive decision, made in a fit of anger, to avoid addressing this root cause (Table~\ref{tab:chatlog:harms} Chat~\raisebox{.5pt}{\textcircled{\raisebox{-.9pt} {6}}}). Encouraging avoidance of the underlying issue could potentially perpetuate or worsen the eating disorder~\cite{garner1997handbook}. This example highlights a critical flaw in \system{}’s ability to provide holistic support, underscoring the need for improved context awareness.

\system{} sometimes struggled to grasp the nuanced aspects of ED contexts mainly because of the absence of meticulous awareness. For example, \system{} often \textbf{encouraged weight-centric focus}. For example, weight-centric focus in Table~\ref{tab:chatlog:harms} Chat~\raisebox{.5pt}{\textcircled{\raisebox{-.9pt} {7}}} could reinforce the idea that weight loss is inherently positive and primary indicator of success. For individuals with ED, this emphasis on weight can perpetuate unhealthy preoccupations with weight and body image~\cite{polivy2004sociocultural}. This weight-centric focus mirrors the harmful advice given by Tessa, the chatbot deployed and later withdrawn by NEDA, which similarly reinforced a weight-loss focus~\cite{neda_harmful_advice_newyorktimes, neda_harmful_advice_cnn}.

Furthermore, \system{} often emphasized the calorie of food, especially \textbf{highlighting low-calorie food} (Table~\ref{tab:chatlog:harms} Chat~\raisebox{.5pt}{\textcircled{\raisebox{-.9pt} {8}}}). This focus can underpin an unhealthy obsession with calories and the misconception that lower-calorie foods are inherently better, which can bolster restrictive eating behaviors~\cite{fairburn2011eating}. It also often \textbf{moralized food choices}. For instance, the phrase in Table~\ref{tab:chatlog:harms} Chat~\raisebox{.5pt}{\textcircled{\raisebox{-.9pt} {8}}}, ``\textit{Except for the cookie, everything you ate this morning is rich in nutrients},'' implies a judgment that the cookie is a less acceptable or \emph{bad} choice. Such language can contribute to a dichotomous view of food (good vs. bad), which is common among individuals with ED and can exacerbate feelings of guilt, shame, or anxiety around eating certain foods~\cite{tylka2013intuitive}.

Moreover, \system{} often praised and \textbf{encouraged unhealthy eating behaviors} for individuals with ED. For example, it suggested picky eating to users (Table~\ref{tab:chatlog:harms} Chat~\raisebox{.5pt}{\textcircled{\raisebox{-.9pt} {9}}}) that can worsen restrictive or unbalanced eating habits. Similarly, it praised users for restricting food intake, a harmful behavior central to anorexia (Table~\ref{tab:chatlog:harms} Chat~\raisebox{.5pt}{\textcircled{{\scriptsize10}}}). Such encouragement or validation can further exacerbate ED.

In addition, \system{} sometimes \textbf{offered unsupported advice}. For instance, it hallucinated `extreme hunger' as a viable strategy, although it is not a strategy but a phenomenon that some people with ED experience after severe food restriction (Table~\ref{tab:chatlog:harms} Chat~\raisebox{.5pt}{\textcircled{{\scriptsize11}}})~\cite{howtodealwithextremehunger, extremehungerinedrecovery}. This could mislead the user into adopting an unsafe approach, \hr{reaffirming the potential risks of harmful advice highlighted in prior research on the safety of LLM-based chatbots~\cite{de2024chatbots, khawaja2023your, denecke2021artificial}.} 


\subsubsection{Users' Unquestioning Belief from Half-Baked Knowledge}

Although \system{} generated various harmful responses, it is noteworthy that none of the participants reported any perceived or experienced detrimental impacts of \system{} during the interviews. This lack of awareness was partly attributed to participants' strong trust in LLM-based chatbots, as observed in both our \hr{results and prior research on AI chatbots in the health domain~\cite{denecke2021artificial, khawaja2023your}.} 
Most participants exhibited strong trust in \system{}, primarily due to their \hr{partial} understanding of how artificial intelligence (AI) operates. Many believe it is inherently trustworthy just because AI is data-driven without fully understanding how AI learns from data. Most considered the training data to be a credible source, which they felt made \system{}'s responses reliable for their ED recovery. For example, P3 said, ``\textit{I thought as AI collects a lot of information and talks based on that, it must include information from people who have experienced and overcome eating disorders and sharing it with me. So, I believed it would contain valuable information for overcoming it}.''

This trust in \system{} sometimes led participants to forgo seeking information from other sources or critically evaluating the information provided by \system{}. P19 noted, ``\textit{Before using this chatbot, I mostly relied on Google for information on symptoms and nutrient deficiencies. However, \system{} offers more comprehensive and contextually relevant information and advice, which I found more helpful than Googling, leading me to depend less on Google.}''
Furthermore, some participants expressed a blind trust in \system{} merely due to recent advancements in AI. P12 remarked, ``\textit{With recent developments in AI,  I have a high confidence level in its information.}''
Despite warnings about the potential for hallucinated responses we provided before the deployment study, no participants questioned the credibility of \system{}'s answer during conversations or post-interviews, indicating an unquestioning trust in the system. 

In summary, we have identified a discrepancy between perceived and actual risks: users tend to trust \system{} due to a limited understanding of how AI functions, but \system{}'s error-prone nature can lead to undesirable responses that may exacerbate ED symptoms.

\subsection{Impact of \system{} on ED Condition Outcomes}

We analyzed the Brief-IPQ survey responses before and after interacting with \system{} to assess the changes in our participants' attitudes toward their ED. The Wilcoxon Signed-Rank Test revealed a statistically significant decrease from pre- to post-interactions ($Z=2.43$, $p=.02$, \hr{$r=0.41$), indicating a medium-to-large effect.} This overall suggests that participants perceived their condition as less severe and felt more in control of managing their illness after interacting with \system{} for their ED recovery. 

\hr{In detail, the Wilcoxon Signed-Rank Test on individual items of the Brief-IPQ showed a statistically significant increase in the \textit{Treatment Control} score ($Z=2.41$, $p=.02$, $r=0.55$), indicating a large effect size, and a significant decrease in the \textit{Concern} score ($Z=1.90$, $p=.05$, $r=0.49$), indicating a medium-to-large effect. No significant differences were found in other items. The increased \textit{Treatment Control} score suggests that participants felt more confident in the effectiveness of their treatment, encompassing \system{} intervention, in helping them manage their illness. The decreased \textit{Concern} score suggests that the intervention helped alleviate participants' concerns about their condition. Each participant's Brief-IPQ survey responses collected both before and after interacting with \system{} are included in our Supplementary Materials.}

We also asked participants in the post-survey, ``Based on your experiences over the past 10 days, how helpful do you think the interactions with the \system{} have been in your eating disorder recovery?'' using a five-point rating (1: Not helpful at all, 2: Not helpful, 3: Neutral, 4: Helpful, 5: Very helpful) to assess the perceived effectiveness of \system{} in ED recovery. Twenty participants rated it as Helpful or Very Helpful, four rated it as Neutral, and one rated it as Not Helpful. The participant who rated it as Not Helpful clarified in a follow-up question that it was not detrimental but simply not helpful. 

%% file: 07_Discussion.tex
\section{Discussion}

We found in our study that many participants with eating disorders felt empowered to continue their recovery journey by discussing their ED-related experiences with an LLM-based chatbot. However, we also identified concerns regarding the users' trust in \system{}'s reliability and various potential harms that could arise due to \system{}'s lack of deep understanding of ED contexts. Building on these findings, we present design implications for LLM-based tools to support users' storytelling practice for mental health, human-LLM collaborative mental health interventions, and encouraging meticulous thinking during human-chatbot interactions.

\subsection{Empowering Mental Health Management and Recovery Through Storytelling}



Previous studies highlighted the importance of sharing personal stories in managing health conditions, including the journey of overcoming one's mental health challenges as our participants did with \system{}~\cite{mamykina2010constructing, adler2012living, nurser2018personal}. In our study, participants strengthened their commitment to recovery from storytelling their recovery journey to \system{}. Through this storytelling, \system{} engaged users on multiple levels: it listened, empathized, and offered advice based on their narratives.

We found that \system{} served as a \emph{personal yet social space}, offering a unique combination of privacy and interaction that distinguishes it from traditional in-person and online social support networks where individuals sometimes hesitate to share mental health issues for fear of judgment~\cite{prizeman2023effects}. Storytelling enables individuals to externalize difficult internal experiences, such as ED-related feelings and thoughts, making their emotional burden more manageable~\cite{nurser2018personal} and relieved~\cite{mamykina2010constructing}. Our participants also exhibited lower Brief-IPQ scores after interacting with \system{}, which implies positive attitude changes towards managing their ED, possibly influenced by the storytelling practice~\cite{mamykina2010constructing}. This highlights the potential of a supportive, private, and interactive space for storytelling ED-related experiences.

We suggest further research explore how to design such personal yet social spaces powered by LLMs to facilitate recovery narratives for individuals with ED or other mental health issues. For instance, the factors contributing to the positive attitude changes in our study might be multi-faceted and not limited to storytelling alone. Therefore, investigating how storytelling with an LLM-based chatbot specifically influences attitude changes in users' health management could provide valuable insights into designing effective interventions that harness storytelling for health improvements.

Storytelling also allows individuals to reaffirm their competence in managing their health condition and maintain a sense of control over it, which is crucial when dealing with mental health issues~\cite{mamykina2010constructing}. 
In our study, participants enhanced their sense of control over ED specifically by sharing and recognizing their recovery \emph{accomplishments} through the process of storytelling with \system{}. \system{} further facilitated users to share accomplishments by prompting them to reflect on their recovery efforts with goal-oriented nudges and positive feedback on shared accomplishments. Focusing on these accomplishments can create a safe and validating environment, conducive to openly discussing mental illness~\cite{glassman2013motivational}.  
Recent studies proposed LLM-based prompting to encourage self-disclosure and self-reflection for promoting mental wellbeing~\cite{kim2024mindfuldiary, kumar2024supporting, nepal2024contextual, lee2020hear}. We suggest further research on tailoring LLM-generated prompts to foster accomplishment-focused storytelling and reflection in addition to mental health challenge-focused prompts. 


The presence of a mental health issue can disrupt how someone sees their life story and even affect their sense of \emph{identity}~\cite{nurser2018personal}. Storytelling has been suggested as a way for people to bring back a sense of order to their lives~\cite{crossley2000narrative}. By telling their story, individuals can reconnect different parts of their identity, linking their past, their roles in society, and their values. The CHI community has explored how storytelling helps individuals construct and maintain their identity in the health domain~\cite{kim2018understanding, mamykina2010constructing, matthews2011my}. In particular, a broad narrative that encompasses both experiences before and after diagnosis helps individuals maintain a consistent sense of self, fostering a positive self-concept during treatment~\cite{mamykina2010constructing, mcadams1993stories, adler2012living}. In our study, we did not observe significant impacts of storytelling on participants' identities, as their narratives primarily focused on their current or recent ED conditions. We suggest further research to explore how LLM-based tools for mental health management could encourage users to reflect on their experiences holistically from a broader, more comprehensive perspective to support the construction of a positive self-identity.

Our study found that \system{} enabled participants to openly share lived experiences related to their ED. Recent studies demonstrated that LLM-based tools could help clinicians better understand their patients' lives outside clinical settings, bridging the gap between psychiatric patients and clinicians~\cite{kim2024mindfuldiary}. 
Similarly, we propose that LLM tools designed for storytelling could offer valuable insights into the experiences of individuals with ED to clinicians, even without a direct clinician-patient relationship. In the context of ED, patients' digital storytelling has already been used to educate healthcare providers~\cite{lamarre2021healthcare}. Building on this, researchers could further study how LLM-based tools can be designed not only to facilitate users' storytelling but also to effectively convey the lived experiences of individuals with ED to clinicians for educational purposes, enhancing their understanding and improving care strategies. We believe such storytelling could offer opportunities to connect patients' personal experiences with clinical care, and we discuss related design opportunities in the following section.

\subsection{Towards Human-LLM Collaborative Mental Health Intervention}

In our study, we identified harmful chatbot responses when addressing ED-related topics. Given the sensitive and nuanced nature of ED treatment, we propose that LLM-based tools be integrated into clinical care as assistive tools rather than stand-alone interventions. Social media platforms employed digital interventions that detect posts suggesting mental health issues~\cite{chancellor2016quantifying, chancellor2016thyghgapp, de2013predicting} and guide users who made or encountered those posts toward professional clinical treatments~\cite{youtube2023ed,youtube2021you,tiktok2021ed}. Likewise, we identified opportunities for close collaboration between clinicians and LLM-based tools as our chatbot detected patients' early warning signs, which could indicate the need for clinical interventions. These tools could act as moderators between clinicians and patients by detecting and flagging important signals in patient interactions, redirecting them to clinicians, similar to human moderators in online mental health support communities~\cite{huh2015clinical}. These flagged insights could be brought to the clinician's attention during clinical sessions, ensuring that critical information is not overlooked.

\system{} helped participants, who often experience a decline in commitment to recovery over time, stay focused on their recovery through goal-oriented nudges based on their Wellness Plan. A similar goal-oriented approach~\cite{grey2018goal} could be applied in collaboration with clinicians. For example, clinicians and patients could establish short-term goals or coping strategies to work on between sessions. During these intervals, LLM-based tools could send reminders and encouragement to help users adhere to their recovery goals. We believe that such approaches could function as essential tools in managing the boundaries between patient and provider interactions, serving as what is known as boundary negotiating artifacts~\cite{chung2016boundary}. 



Despite the potential of LLM-based tools, the risks of unsupervised interactions remain. To mitigate this, we suggest designing specialized LLMs tailored specifically to the context of ED and mental health care, as generic LLMs may not fully address the complexities of these conditions, as shown in our study~\cite{johnson2023assessing}. Specialized models could offer safer, more contextually appropriate responses~\cite{kraljevic2021medgpt}. Additionally, principled metrics are needed to rigorously evaluate the safety and effectiveness of these LLMs~\cite{johnson2023assessing}, ensuring they provide meaningful support. However, human intervention remains crucial; clinicians should guide these interactions, monitor for risks, and intervene when necessary to ensure patient safety.


In our study, \system{} encouraged a participant to seek support from those around her, which led her family to actively participate in her recovery. Building on this, we suggest further exploration into how family members, patients, and LLM-based chatbots can collaborate to enhance the recovery journey. In ED treatments, family-based therapy is more effective over time than individual therapy, particularly as treatment duration increases~\cite{couturier2013efficacy}. A recent study presenting a framework for designing family informatics tools~\cite{lukoff2018tablechat} suggested that technology should complement, not replace, the emotional support provided by family members, encouraging them to stay engaged and follow up. We propose that chatbots could act as facilitators between patients and their families, supporting long-term recovery by fostering ongoing family involvement in managing eating disorders.


\subsection{Encourage Critical Thinking in Human-Chatbot Interaction}

Our participants exhibited strong trust in \system{}'s reliability, leading to a lack of critical assessment of its responses. This blind trust can pose significant risks, particularly in high-stakes mental health contexts, where inaccurate or inappropriate advice may go unquestioned, as observed in our study. Despite informing participants multiple times about the potential for harmful chatbot advice and encouraging them to critically evaluate responses, none reported identifying any such misinformation or harmful advice.

A key reason for such strong trust was that participants partially understood how LLM-based chatbots work. This gave them enough confidence but not enough awareness to recognize its limitations. \hr{This observation highlights the role of algorithmic understanding in shaping user trust in LLMs. Prior research has emphasized the importance of educating users about LLM functionality to build trust~\cite{zhao2024assessing, schwartz2023enhancing, huschens2023you}. However, our finding that participants exhibited strong trust despite limited understanding underscores the need for deeper investigations into how LLM transparency affects trust. Such research is essential to develop transparency strategies that foster appropriate trust levels~\cite{liao2023ai}.
}

To address \hr{the lack of critical assessment during interactions with chatbots,} we recommend incorporating \emph{in-situ} interventions to encourage users to assess chatbot responses, improving the safety and awareness of these interactions. A 
recent research showed that \emph{in-place} interventions, such as highlighting content that might be wrong or debatable~\cite{hartwig2024adolescents}, raised users' awareness of potential misinformation and encouraged them to critically assess and moderate content~\cite{jahanbakhsh2024browser}.

Taking this further, along with our observation, we suggest that educational interventions
should occur during chatbot use in real-time to promote awareness and critical assessment, as our pre-use educational interventions did not effectively promote critical assessment. 
For instance, while some off-the-shelf chatbots use alerts to inform the fallibility of LLM-based chatbots to users during interaction (e.g., `ChatGPT can make mistakes. Check important info.')~\cite{apple2024chatgpt}, these could be enhanced by explaining why the responses might be inaccurate (e.g., `ChatGPT can make mistakes \emph{as it generates responses from patterns in training data, so check its responses.}') to help users better understand LLM-based chatbots' limitations and inform users to assess responses more carefully. However, this transparency should be carefully balanced to avoid overwhelming users with excessive details, which could erode trust~\cite{kizilcec2016much}, while preventing chatbot responses' uncritical acceptance. Further research is needed to explore how to optimize the amount and type of transparency provided during chatbot interactions, ensuring it reduces the risks of over-reliance on incomplete chatbot responses without undermining user trust~\cite{liao2023ai}.

We found that participants in our study not only failed to critically assess the \system{}'s responses but also that \system{} itself occasionally provided careless responses. To address LLM unreliability, recent studies explored incorporating human cognitive practices such as self-reflection into LLMs, leading to performance improvements in certain domains~\cite{mcaleese2024llm, ji2023towards, renze2024self}. However, the outcomes of applying different human cognitive practices to LLMs varied across different applications, including the medical domain. For instance, the \textit{retry} self-reflection strategy performed best in math problem-solving but worst in analytic reasoning tasks within the legal domain~\cite{renze2024self}. In programming code review, introducing criticism sometimes caused the model to hallucinate non-existent problems~\cite{mcaleese2024llm}. We recommend further research to carefully examine how these strategies impact LLM performance in various mental health contexts. We also propose exploring design approaches that encourage users to engage the LLM in effective cognitive practices during their interactions with LLMs. We believe that a deeper investigation into mindful user interactions with LLM-based chatbots could yield valuable insights that enhance user safety and inform broader applications of LLMs. 



\subsection{Limitations}
The 10-day study period may have constrained our findings. The longer deployment study could reveal additional, unique benefits and harms from extended interactions with LLM-based chatbots. In addition, our participant pool was skewed toward females, which reflects the higher prevalence of eating disorders among women than men~\cite{galmiche2019prevalence, kessler2013prevalence}. However, a more diverse gender representation would have been beneficial to explore how LLM-based chatbots impact individuals across a broader spectrum of gender identities. 
\hr{Our study was conducted on a single LLM-based chatbot. Incorporating additional chatbot baselines using different LLMs could help uncover model-specific variations in chatbot responses. Lastly, we did not perform an ablation study, which could also provide deeper insights into how each feature specifically contributes to user experience and intervention effectiveness. We believe these could be valuable directions for future research.}

%% file: 08_Conclusion.tex
\section{Conclusion}
Chatbots have been recognized as effective digital interventions for individuals with eating disorders. Amid the growing sophistication of LLMs, LLM-based chatbots are increasingly being deployed, and people with various mental health issues, including eating disorders, are turning to them for support. However, in high-stakes mental health contexts, their use raises concerns due to their fallibility, such as the potential for inaccurate or harmful responses. In this study, we examined the benefits and shortcomings of using LLM chatbots for everyday eating disorder care. We found that \system{} provided \emph{private yet social} space, a unique combination of privacy and interaction distinct from traditional support networks, in which users could freely discuss their eating disorders without social stigma and receive emotional and informational support. We also identified risks stemming from \system{}’s occasional harmful and careless responses and users’ strong trust in \system{}’s guidance without critical evaluation. Building on these findings, we share insights for designing LLM-powered digital interventions for eating disorders and other mental health issues.
